\documentclass[twocolumn,secnumarabic,amssymb,amsmath,showpacs,nobibnotes,aps,prb]{revtex4-1}

\usepackage[left=1in,right=1in,top=1in,bottom=1in]{geometry}

\usepackage{tabularx}
\usepackage{float}
\usepackage{bm}

\usepackage{graphicx}
\usepackage{amsmath}
\usepackage{color}

\begin{document}

\title{Optically controlled periodical chain of quantum rings}

\author{M.~Hasan$^{1}$}

\author{I. V.~Iorsh$^{1}$}\email{i.iorsh@phoi.ifmo.ru}
\author{O. V.~Kibis$^{2, 3}$}
\author{I. A.~Shelykh$^{1, 3, 4}$}

\affiliation {$^1$ITMO University, St. Petersburg 197101, Russia}
\affiliation {$^2$Department of Applied and Theoretical Physics,
Novosibirsk State Technical University, Karl Marx Avenue 20,
Novosibirsk 630073, Russia} \affiliation {$^3$Science Institute,
University of Iceland, Dunhagi-3, IS-107 Reykjavik, Iceland}
\affiliation {$^4$Division of Physics and Applied Physics, Nanyang
Technological University 637371, Singapore}
\begin{abstract}
We demonstrated theoretically that a circularly polarized
electromagnetic field substantially modifies electronic properties
of a periodical chain of quantum rings. Particularly, the field
opens band gaps in the electron energy spectrum of the chain,
generates edge electron currents and induces the Fano-like
features in the electron transport through the finite chain. These
effects create physical prerequisites for the development of
optically controlled nanodevices based on a set of coupled quantum
rings.
\end{abstract}


\maketitle

\section{Introduction}
Progress in fabrication of semiconductor nanostructures has led to
achievements in studies of various ring-like mesoscopic objects,
including quantum rings, nanotubes, nanohelices (see, e.g.,
Refs.~[\onlinecite{Fomin,Saito,Prinz_00,Kibis_92}]). The physical
interest to the rings is caused by the phenomenon of the
interference of electron waves, which can be observed there.
Particularly, the Aharonov-Bohm (AB) effect arisen from the direct
influence of the vector potential on the phase of the electron
wave
function~\cite{AB_Chambers1960,ABoriginal,AB_prallel_1984,AB_topology}
has been studied both theoretically and experimentally in various
ring-like
nanostructures~\cite{AB_ballistic_exp_1,AB_ballistic_exp_2,AB_ballistic_exp_3,AB_ballistic_exp_4,AAS_exp1,AASExp2,
AB_WeakLoc2,AB_WeakLoc3,AB_bal_WL_2005}. Conceptually, the AB
effect is caused by the broken time-reversal symmetry in an
electron system  subjected to a magnetic flux. Namely, the flux
breaks the physical equivalence of clockwise and counterclockwise
electron rotations in a ring, which results in the flux-controlled
interference of the electron waves corresponding to these
rotations. However, the time-reversal symmetry can be broken not
only by a magnetic flux but also by a circularly polarized
electromagnetic field. Therefore, the strong coupling of electrons
in quantum rings to off-resonant circularly polarized photons
leads to the optically induced AB
effect~\cite{dressing_kibisprl,Kibis_14,Shelykh2014AB_opt,exciton_kibis}.
As a consequence, stationary electronic properties of quantum
rings can be effectively controlled with light. It should be noted
that the optical control of quantum rings is attractive from
applied viewpoint since it is much faster than the
magnetic-flux-induced control. Therefore, optically-controlled
ring-like nanostructures can be considered as a basis for creating
ultra-fast logic gates. In the previous studies, the main
attention was paid to the optically induced effects in sole
quantum rings. As to the effects in multi-ring systems, they
escaped attention before. In the present article, we perform
theoretical analysis of an one-dimensional chain of coupled
quantum rings~\cite{QRchain00,Chains11,ChainsFib,Chains7}
subjected to an off-resonant circularly polarized electromagnetic
wave and demonstrate that electronic properties of the chain are
very sensitive to the irradiation.

\section{Model}
\begin{figure}[!h]
\centerline{\includegraphics[width =
1.0\columnwidth]{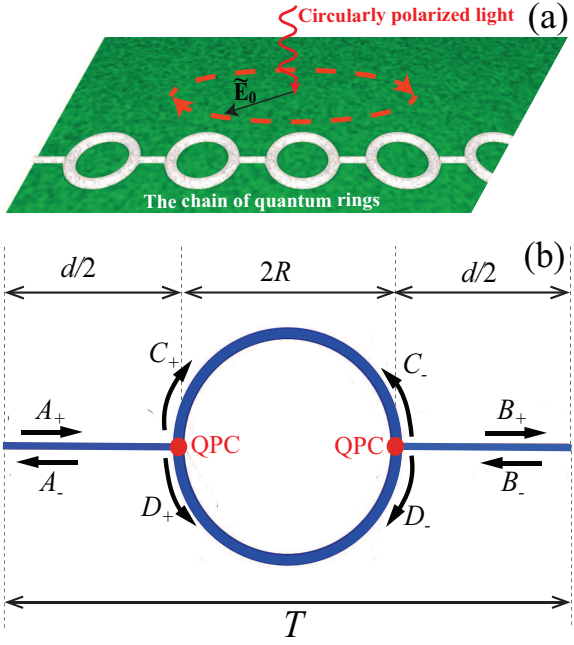}} \caption{Sketch of the system
under consideration: (a) The periodical chain of quantum rings
irradiated by a circularly polarized electromagnetic wave with the
electric field amplitude $\widetilde{E}_0$; (b) The elementary
cell of the chain consisting of a quantum ring with the radius
$R$, two leads with the length $d/2$ and two quantum point
contacts (QPC). The arrows correspond to electron waves traveling
in different ways with the amplitudes $A_\pm$, $B_\pm$, $C_\pm$,
$D_\pm$, and $T$ is the period of the chain} \label{fig_1}
\end{figure}

Let us consider the periodical chain of quantum rings irradiated
by a circularly polarized electromagnetic wave with the electric
field amplitude $\widetilde{E}_0$ and the frequency $\omega$,
which is assumed to be far from resonant frequencies of the
electron system (see Fig.~\ref{fig_1}).
Within the scattering matrix
approach~\cite{AB_bal_WL_2005,Butticker1984}, the amplitudes of
electron waves in the chain, $A_\pm$, $B_\pm$, $C_\pm$, $D_\pm$,
satisfy two equations,
\begin{equation}
\left(
\begin{array}{c}
 A_- e^{{-iqd}/{2}} \\
 C_+ \\
 D_+ \\
\end{array}
\right)=S \left(
\begin{array}{c}
 A_+ e^{{i qd}/{2}} \\
 C_- e^{i \left(\pi  q R-\phi _0\right)} \\
 D_- e^{i \left(\pi  q R+\phi _0\right)} \\
\end{array}
\right), \label{eq:leftQPC}
\end{equation}
\begin{equation}
\left(
\begin{array}{c}
 B_+ e^{-i qd/2} \\
 C_- \\
 D_- \\
\end{array}
\right)=S \left(
\begin{array}{c}
 B_- e^{{iqd }/{2}} \\
 C_+ e^{i \left(\pi  q R+\phi _0\right)} \\
 D_+ e^{i \left(\pi  q R-\phi _0\right)} \\
\end{array}
\right), \label{eq:rightQPC}
\end{equation}
where the scattering matrix is
\begin{equation}
S=\left(
\begin{array}{ccc}
 \sqrt{1-2\varepsilon^2}  & \varepsilon  & \varepsilon  \\
 \varepsilon  & \frac{-(1+\sqrt{1-2\varepsilon^2})}{2} & \frac{(1-\sqrt{1-2\varepsilon^2})}{2} \\
 \varepsilon  & \frac{(1-\sqrt{1-2\varepsilon^2})}{2} & \frac{-(1+\sqrt{1-2\varepsilon^2})}{2} \\
\end{array}
\right), \label{eq:S1}
\end{equation}
$\varepsilon$ is the electron transmission amplitude through the
QPCs ($0\leq\varepsilon\leq1/\sqrt{2}$), $q=\sqrt{2m_e E/\hbar^2}$
is the electron wavenumber, $m_e$ is the electron mass, and $E$ is
the electron energy. As to the phase shift in
Eqs.~(\ref{eq:leftQPC}) and (\ref{eq:rightQPC}),
\begin{align}
\phi_0=\frac{\pi e^2\widetilde{E}_0^2}{m_e\hbar \omega^3},
\label{eq:phi0}
\end{align}
it describes the phase difference for electron waves traveling
inside the ring clockwisely and counterclockwisely, which arises
from the electron coupling to the circularly polarized
irradiation~\cite{Shelykh2014AB_opt}. It should be stressed that
the phase shift (\ref{eq:phi0}) is induced by an off-resonant
electromagnetic field (``dressing field'' in conventional terms of
quantum optics) which cannot be absorbed by electrons. Therefore,
the effects discussed below substantially differ from the effects
caused by absorption of light in quantum rings (see, e.g.,
Refs.~[\onlinecite{Pershin_2005,Rasanen_2007,Matos_2005}]).

\begin{figure}[!h]
\centerline{\includegraphics[width = 1.0\columnwidth]{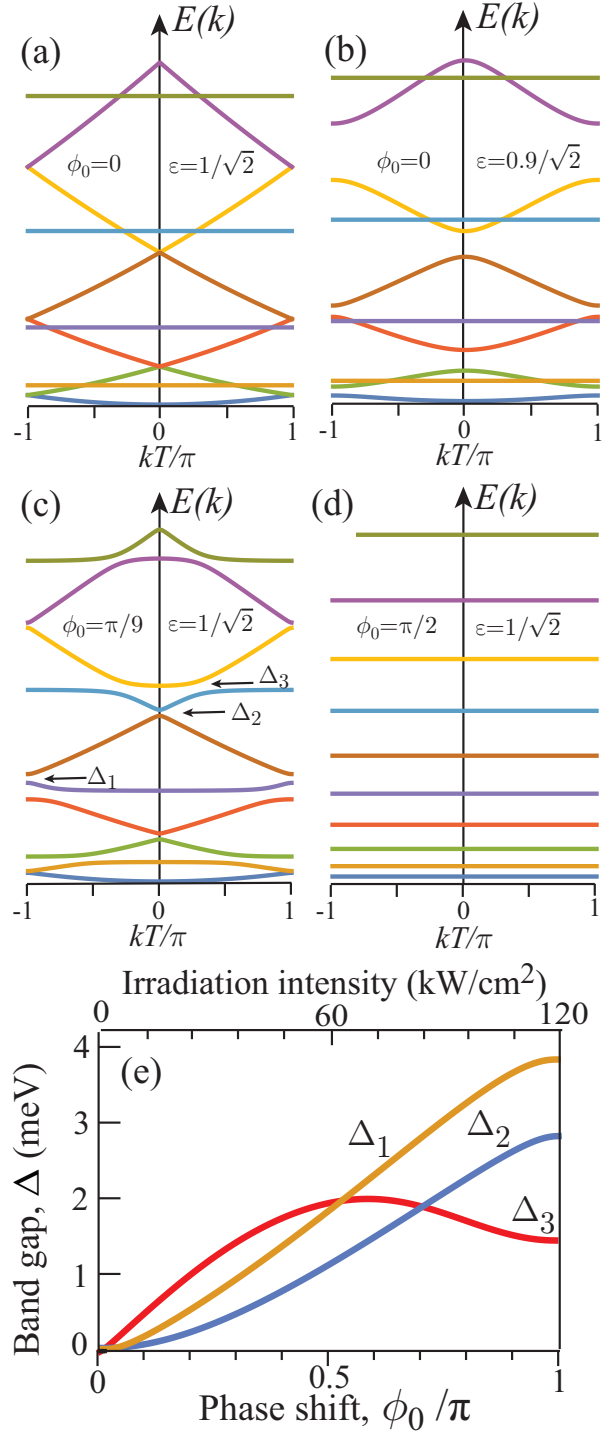}}
\caption{Electron energy spectrum of the chain of GaAs-based
quantum rings with the radius $R=30$~nm and the period $T=100$~nm
in the presence of a circularly polarized electromagnetic field
with the frequency $\omega=12\times 10^{12}$ rad$/$s.}
\label{fig_2}
\end{figure}

Applying the Bloch theorem to the considered periodic chain of
quantum rings, we arrive at the equation
\begin{equation}
\left(
\begin{array}{c}
 A_+ \\
 A_- \\
\end{array}
\right)=e^{i k T}\left(
\begin{array}{c}
 B_+ \\
 B_- \\
\end{array}
\right) , \label{eq:Bloch}
\end{equation}
where $k$ is the electron wave vector originated from the
periodicity of the chain and $T=d+2R$ is the period of the chain.
Mathematically, Eqs.~(\ref{eq:leftQPC}), (\ref{eq:rightQPC}) and
(\ref{eq:Bloch}) form a homogeneous system of linear algebraic
equations for the eight amplitudes
$A_{\pm},B_{\pm},C_{\pm},D_{\pm}$. The secular equation of the
algebraic system,
\begin{equation}
\begin{aligned}
&\sin(qd)\Big[(1-\varepsilon^2)\cos^2\phi_0+\sqrt{1-2 \varepsilon^2}\sin^2\phi_0\\
&-\cos(2\pi q R)\Big] +\varepsilon^2\Big[ \sin[q(d-2\pi R)]\\
&+ 2 \cos\phi_0\sin(\pi  q  R)\cos(k T)\Big]=0,
\end{aligned}
\label{eq:dispersion}
\end{equation}
defines the sought electron energy spectrum of the irradiated
chain, $E(k)$, which is plotted in Fig.~\ref{fig_2} for the
particular important cases discussed below.

\section{Discussion and conclusions} The electron energy spectrum of the chain
without the irradiation ($\phi_0=0$) is shown in
Figs.~\ref{fig_2}(a)--(b). In the case of transparent QPCs
($\varepsilon=1/\sqrt{2}$),  Eq.~(\ref{eq:dispersion}) reads as
\begin{equation}
\sin(q\pi R)\left[\cos(kT)-\cos(qd+q\pi R)\right]=0.
\label{eq:Deg}
\end{equation}
Evidently, Eq.~(\ref{eq:Deg}) has the two solutions, $q=(|kT| +
2\pi n)/(d+\pi R)$ and $q=m/R$, where $n,m=0,\pm1,\pm2,...$. The
first solution produces the $k$-dependent branches of the electron
energy spectrum, $E(k)$, which correspond to an electron
propagating along the chain, whereas the second one results in
flat bands corresponding to an electron localized inside
individual rings [see Fig.~\ref{fig_2}(a)]. In the case of
semitransparent QPCs ($0<\varepsilon<1/\sqrt{2}$), the secular
equation (\ref{eq:dispersion}) takes the form
\begin{align}\label{st}
&\Big[( 2 \varepsilon^2 - 1) \cos[q (d - \pi R)] + \cos[q (d + \pi
R)]
\nonumber \\
&-  2 \varepsilon^2 \cos(k T)\Big]\times \sin(q \pi R) = 0.
\end{align}
It follows from Eq.~(\ref{st}) that the non-transparency of the
QPCs leads to opening band gaps at the Brillouin zone edges
($k=\pm\pi/T$), which arise from the Bragg reflection of electron
waves by the QPCs [see Fig.~\ref{fig_2}(b)]. As to the limit of
weakly transparent QPCs, $\varepsilon \ll 1$, the band structure
contains only the electron modes localized inside rings and leads,
which are defined by the dispersion equation $\sin (qd)\sin (q\pi
R)=0$.

The electron energy spectrum of the irradiated chain
($\phi_0\neq0$) is shown in Figs.~\ref{fig_2}(c)--(d). Since the
localized electron modes are not eigenmodes of the irradiated
structure, the phase shift $\phi_0\neq0$ results in the coupling
between the localized electron modes of the rings and the
propagating electron modes of the chain. As a consequence, the
anticrossings --- which manifests itself through opening of the
additional band gaps inside the Brillouin zone --- appears [see
Fig.~\ref{fig_2}(c)]. It should be noted also that in the
particular case of $\phi_0=\pi/2$ Eq.~\eqref{eq:dispersion}
results in the series of the $k$-independent solutions producing
flat bands in Fig.~\ref{fig_2}(d). Thus, generally, the two types
of the band gaps can be identified in the considered chain: The
type-I gaps can take place in the unirradiated chain, whereas the
type-II gaps are caused by the irradiation. Namely, the type-I
gaps, $\Delta_1$ and $\Delta_2$, are opened at edges and at the
center of the Brillouin zone, whereas the type-II gaps,
$\Delta_3$, appear at the crossing of delocalized and localized
electron modes [see Fig.~\ref{fig_2}(c)]. It should be noted that
these two types of the band gaps have different dependence on the
small phase shifts $\phi_0$: While the type-I gaps are
$\Delta_{1,2}\propto \phi_0^2$, the type-II gap is
$\Delta_{3}\propto \phi_0$. The explicit forms of the asymptotic
expressions for the gaps at $\phi_0\ll1$ read as
\begin{align}
&\Delta_1\approx \frac{3\pi\hbar^2}{m_e  (d+\pi R)^2}\cot\left(\frac{3\pi}{2}\frac{\pi R}{d+\pi R}\right)\phi_0^2,\\
&\Delta_2\approx \frac{4\pi\hbar^2}{m_e  (d+\pi R)^2}\tan\left(2\pi\frac{\pi R}{d+\pi R}\right)\phi_0^2,\\
&\Delta_3\approx \frac{3\hbar^2}{m_e R^2}\sqrt{\frac{2R}{\pi(d+\pi
R)}}\phi_0
\end{align}
The dependence of gaps $\Delta_{1,2,3}$ on the phase shift
$\phi_0$ is plotted in Fig.~\ref{fig_2}(e). The
irradiation-induced type-II gaps are most interesting from
viewpoint of possible applications since they allow to design the
optically-controlled modulators of the electron signal propagating
in the chain.
\begin{figure}[!h]
\centerline{\includegraphics[width =
1.0\columnwidth]{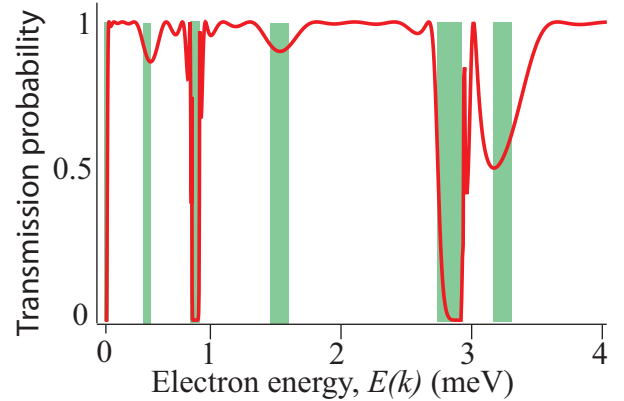}} \caption{Probability of electron
transmission through a finite chain consisting of five quantum
rings at the irradiation-induced phase shift $\phi_0=\pi/9$. The
shaded-green regions correspond to the band gaps in
Fig.~\ref{fig_2}(c).}\label{fig_3}
\end{figure}
It should be noted that the coupling of the narrow localized
electron mode and the delocalized one, which results in opening
the type-II band gaps, can be described formally in terms of the
Fano resonance~\cite{Fano1961, Fano_Rev}. In order to demonstrate
this, we calculated the probability of electron transmission
through a finite chain of quantum rings (see the Appendix). The
Fano-like asymmetry of the lineshape of the transmission spectrum
is clearly seen in Fig.~\ref{fig_3}.

\begin{figure}[H]
\centerline{\includegraphics[width =
1.0\columnwidth]{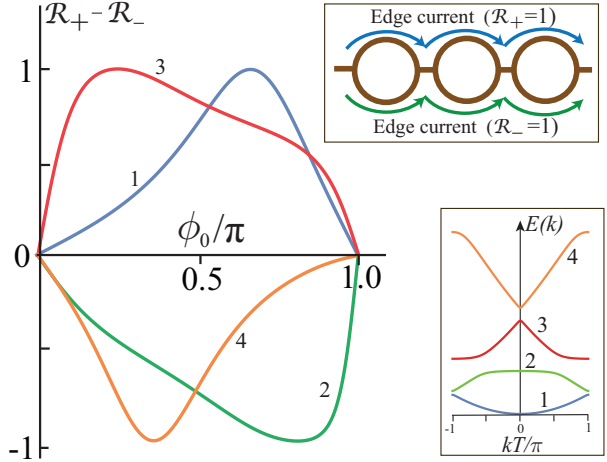}} \caption{Distribution of electron
density in the chain, $\mathcal{R}=\mathcal{R}_+-\mathcal{R}_-$,
as a function of the irradiation-induced phase shift $\phi_0$ at
$kT = \pi/4$ for the four lowest energy bands pictured
schematically in the lower insert, where $\mathcal{R}_+$ and
$\mathcal{R}_-$ are the probabilities of electron stay in upper
and lower segments of the rings, respectively. The upper insert
shows the edge currents corresponding to the cases of
$\mathcal{R}_\pm=1$.} \label{fig_4}
\end{figure}

In order to demonstrate another interesting irradiation-induced
effect in the considered system, let us introduce the
probabilities to find the electron in upper and lower segments of
rings, $\mathcal{R}_\pm$. The difference of these probabilities,
$\mathcal{R}_+-\mathcal{R}_-$, which gives the distribution of
electron density in the chain, is plotted in Fig.~\ref{fig_4} as a
function of the phase shift, $\phi_0$. It follows from the plot
that an electron in the irradiated chain propagates preferably
either in the upper or the lower segments of the rings. Moreover,
one can see that $\mathcal{R}=\pm1$ for some values of $\phi_0$.
As a consequence, electrons propagating along the chain can be
localized in different (upper or lower) segments of the rings.
Thus, the irradiation-induced edge currents appear (see the upper
insert in Fig.~\ref{fig_4}).

In the present paper, we restricted our consideration by the
simplest case of the periodic chain consisting of identical rings.
It is interesting to discuss the more complicated case of the
doubly-periodic chain, where the unit cell of the periodic
structure consists of two non-equivalent quantum rings. In this
case, the simultaneous breakdown of the time-reversal symmetry
(due to the applied circularly polarized field) and the inversion
symmetry (due to non-equivalency of the rings) takes place. It
follows from the fundamentals of quantum mechanics that these
broken symmetries result in the asymmetric electron dispersion,
$E(k)\neq E(-k)$, which has been studied in various nanostructures
(see, e.g.,
Refs.~\onlinecite{Kibis2001_nanotubes,Lawton2002_rectification}
and references therein). As a consequence, electronic phenomena
which are specific for such an asymmetric dispersion can be
expected also in the system under consideration.

Finalizing the discussion, let us analyze observability of the
predicted effects. There are the two fundamental restrictions for
the developed theory. Firstly, the mean free path of electron for
inelastic scatterring processes should be much greater than the
chain period $T$. Secondly, the condition of strong electron-field
coupling must be satisfied: The time of electron travelling
through a ring should be much less than the field period,
$2\pi/\omega$~\cite{Shelykh2014AB_opt}. In the considered case of
GaAs-based quantum rings with the Fermi energy of meV scale and
the chain period $T\sim10^{-5}$~m, the both conditions can be
satisfied for a dressing field of THz frequency range with the
intensity $I_0\sim10^{4}$~W/cm$^2$. It follows from the
calculations that the field-induced band gaps $\Delta_{1,2,3}$ are
of meV scale [see Fig.~\ref{fig_2}(e)] and, therefore, can be
easily detected in state-of-the-art optical experiments. To use
the irradiation frequencies within the visible and near-infrared
ranges, the field intensity, $I_0=\widetilde{E}_0^2c/8\pi$, should
be increased in order to keep the phase shift (\ref{eq:phi0}) to
be large enough. However, increasing the field intensity can melt
the nanostructure. To avoid the melting with the strong field, it
is reasonable to use narrow pulses of a circularly polarized field
which open band gaps and narrow pulses of a weak probing field
which detect the gaps. This pump-and-probe methodology has been
elaborated long ago and is commonly used to observe various
effects induced by strong fields
--- particularly, modifications of electron energy spectrum arisen
from the optical Stark effect
--- in semiconductor structures \cite{Joffre1,Joffre2,PumpProbe1}.
Since giant field intensities (up to GW/cm$^2$) can be applied to
semiconductor structures within this approach, the wide band gaps
can be opened with the pulsing fields.

Summarizing the aforesaid, we analyzed the electron dispersion and
electron transport properties of an one-dimensional periodical
chain of quantum rings irradiated by circularly polarized light.
It is shown that the optically induced Aharonov-Bohm effect leads
to opening band gaps in electron energy spectra of the chain and
modifies electron transport through the chain. These findings can
be exploited, for instance, in optically controlled logic gates
with high operation speed.

The work was partially supported by FP7 IRSES projects POLATER and
QOCaN, FP7 ITN project NOTEDEV, Rannis project BOFEHYSS, RFBR
projects 14-02-00033 and 16-02-01058, the Russian Ministry of
Education and Science, the Russian Target Federal Program
``Research and Development in Priority Areas of Development of the
Russian Scientific and Technological Complex for 2014-2020''
(project 14.587.21.0020)

\appendix
\section{Derivation of the transmission probability for a finite chain of
quantum rings} In order to obtain the expression for the
probability of electron transmission through a finite periodic
chain of quantum rings (the transmission probability plotted in
Fig.~4), we have to introduce the transfer matrix,
$\hat{\mathrm{T}}$, which connects the incoming electron
amplitudes, $A_{\pm}$, with the outgoing electron amplitudes,
$B_{\pm}$,
\begin{align}\label{S2}
\begin{pmatrix} B_+ \\ B_- \end{pmatrix} = \hat{\mathrm{T}}\begin{pmatrix} A_+ \\ A_-
\end{pmatrix}.
\end{align}
It follows from the basic equations (1)--(2) that the transfer
matrix for the elementary cell pictured in Fig.~1(b) reads as
\begin{widetext}
\begin{align}\label{S1}
\hat{\mathrm{T}}=\frac{1}{2\cos {\phi _0}\sin (\pi
qR)}\begin{pmatrix}i {{e^{iqd}}\left[{{\cos }^2}{\phi _0} -
{e^{2iq\pi R}}\right]} & -i { {{\sin }^2}{\phi _0}} \\ i { {{\sin
}^2}{\phi _0}} & -i { {e^{ - iqd}}\left[{{\cos }^2}{\phi _0} -
{e^{ - 2iq\pi R}}\right]}
\end{pmatrix}.
\end{align}
\end{widetext}
The transfer matrix over $N$ elementary cells,
$\hat{\mathrm{T}}^{(N)}$, is just the $N$-th power of the transfer
matrix (\ref{S1}), i.e.
$\hat{\mathrm{T}}^{(N)}=\hat{\mathrm{T}}^{N}$. If the incoming
amplitude is $A_+=1$, we arrive from Eq.~(\ref{S2}) to the
equation
\begin{align}\label{S3}
\hat{\mathrm{T}}^{(N)}\begin{pmatrix} 1 \\
r\end{pmatrix} = \begin{pmatrix} t \\ 0\end{pmatrix},
\end{align}
where $r$ and $t$ are the amplitudes of electron reflection and
electron transmission through the chain, respectively. Taking into
account the unimodularity property of the transfer matrices, the
solving of Eq.~(\ref{S3}) leads to the transmission amplitude
\begin{align}\label{S4}
t=\frac{1}{{\mathrm{T}}^{(N)}_{22}},
\end{align}
where
\begin{eqnarray}\label{S5}
{\mathrm{T}}^{(N)}_{22}&=&{{\mathrm{T}}_{22}}{U_{N -
1}}\left(\frac{{\mathrm{T}}_{11}+{\mathrm{T}}_{22}}{2}\right)\nonumber\\
&-&{U_{N -
2}}\left(\frac{{\mathrm{T}}_{11}+{\mathrm{T}}_{22}}{2}\right),
\end{eqnarray}
and $U_i(x)$ is the Chebyshev polynomial of the second
kind~\cite{Markos2008}. Using Eqs.~(\ref{S4}) and (\ref{S5}), we
can easily calculate the transmission probability, $|t|^2$,
plotted in Fig.~4.

\end{document}